\documentclass{article}
\usepackage{spconf}
\usepackage[pdftex]{graphicx}
\usepackage{subfigure}
\usepackage{epstopdf}
\usepackage{multirow}
\usepackage{url}
\usepackage{amsfonts}
\usepackage{amsmath}
\usepackage{tabularx}

\usepackage{tikz}
\usepackage{textcomp}
\usepackage{lipsum}

\newcommand\copyrighttext{%
  {\Huge IEEE Copyright Notice}
  \\
  \\
  \textcopyright 2012 IEEE. Personal use of this material is permitted. Permission from IEEE must be obtained for all other uses, in any current or future media, including reprinting/republishing this material for advertising or promotional purposes, creating new collective works, for resale or redistribution to servers or lists, or reuse of any copyrighted component of this work in other works. 
  \\
  \\
  \\
  \\
  \\
  \\
  \\
  \\
  \\
  \\
  {\Large Accepted to be Published in: IEEE International Conference on Image Processing, September 22-25, 2019, Taipei, Taiwan}
  }
\newcommand\copyrightnotice{%
\begin{tikzpicture}[remember picture,overlay]
\node[anchor=center,yshift=200pt] at (current page.center)
{\parbox{\dimexpr\textwidth-\fboxsep-\fboxrule\relax}{\copyrighttext}};
\end{tikzpicture}%
}

\begin{document}
%
%
\copyrightnotice
%

\title{Compressed Image Quality Assessment Based on Saak Features}

\name{Xinfeng Zhang$^{\S\dag\ddag}$, Sam Kwong$^{\dag}$ and C.-C. Jay Kuo$^{\ddag}$}
\address{$^{\S}$School of Computer Science and Technology, University of Chinese Academy of Sciences\\
$^{\dag}$Department of Computer Science, City University of Hong Kong\\
$^{\ddag}$Ming Hsieh Department of Electrical Engineering, University of Southern California}

\maketitle

\begin{abstract}
Compressed image quality assessment plays an important role in image
services, especially in image compression applications, which can be
utilized as a guidance to optimize image processing algorithms. In this
paper, we propose an objective image quality assessment algorithm to
measure the quality of compressed images. The proposed method utilizes a
data-driven transform, Saak (Subspace approximation with augmented
kernels), to decompose images into hierarchical structural feature
space. We measure the distortions of Saak features and accumulate these
distortions according to the feature importance to human visual system.
Compared with the state-of-the-art image quality assessment methods on
widely utilized datasets, the proposed method correlates better with the
subjective results. In addition, the proposed methods achieves more
robust results on different datasets.
\end{abstract}
\section{Introduction}
\label{sec:intro}
Lossy image compression techniques such as JPEG and JPEG2000 achieved high compression ratios at the cost of perceived degradation in image quality \cite{zhang2013compression,zhang2016low,zhang2017high}. The state-of-the-art image compression systems usually need a quality metric to optimize the image coding procedure by assigning bits to various image contents adaptively. However, the widely utilized peak-signal-to-noise-ratio (PSNR) metric mainly focuses on the pixel-level difference between the original and compressed images, which is not well correlated with human perceptual quality. This is because human perception is very sensitive to structural distortions instead of individual pixel distortion.

In the past decades, to obtain more consistent quality measures with human visual perception, numerous image quality assessment (IQA) methods \cite{wang2004image,wang2003multiscale,wang2011information,zhang2010rfsim,zhang2011fsim,gore2015full,li2016no} and datasets \cite{LIVEdataset,larson2010most,zhang2019fine} have been proposed for different distortion types. The existing IQA methods can be divided into three categories according to the availability of reference images, \textit{i.e.}, full reference IQA (FR-IQA), reduced reference IQA (RF-IQA) and no reference IQA (NR-IQA). In most scenarios, the reference images are available for compressed image quality assessment problem. In \cite{wang2004image}, Wang \textit{et al.} proposed Structural SIMilarity (SSIM) index metric to calculate image quality according to patch similarity between the reference and distorted images instead of the pixel-level distortions. In addition, SSIM takes the correlation function as its quality metric model to reflect patch structural distortions instead of mean square error (MSE), which is utilized in PSNR. This simple and effective method achieved more correlated results with subjective quality on different IQA datasets.

To further improve the performance of SSIM, in \cite{wang2003multiscale}, Wang \textit{et al.} measured multi-scale structural similarity (MS-SSIM) to provide more flexibility than single-scale method by dealing with the variational perceivability of image details. In \cite{wang2011information}, Wang and Li further utilized the information content weighting strategy to improve the IQA performance in pooling stage, where the weights are calculated based on the Gaussian scale mixture (GSM) model of natural images. This weighting strategy not only can improve the performance of SSIM metric but also can make the pixel-level quality metric PSNR be a competitive perceptual quality metric. In \cite{zhang2010rfsim,zhang2011fsim}, Zhang \textit{et al.} further improved the performance of the SSIM-based quality metric by designing more effective features to capture the local structural information, to which human visual system (HVS) is more sensitive than local pixel correlation in SSIM.

Different from other distortion types, the major distortions in compressed images are structural distortions, \textit{e.g.} blocking artifacts in JPEG images and ringing artifacts in JPEG2000 images. Therefore, highly efficient structural information representation is more important for compressed image quality assessment problem. In \cite{gore2015full}, Gore and Gupta proposed the LSDBIQ technique by utilizing the local standard deviation of an image to reflect the structural information to evaluate the compressed JPEG image quality. In \cite{li2016no}, Li \textit{et al.} evaluated the structural distortions including blockiness and blurring artifacts by utilizing the Tchebichef moments as features to differentiate the quality of deblocked JPEG compression images.

In this paper, we proposed a new full reference image quality assessment method based on the efficient data-driven feature extraction transform, Subspace approximation with augmented Kernels (Saak) \cite{kuo2018data}. The proposed method first applies a low-pass filter to remove random noise which is difficult to be perceived by HVS. Then, the data-driven Saak transform is learnt from reference image, and both reference and distorted images are transformed by the learnt Saak transform, which converts images from pixel domain to feature domain. Since the feature maps in different spectral components of Saak transform domain show various influences on perceptual quality, we design a weighting strategy for the qualities of feature maps. The experimental results on JPEG and JPEG2000 images are conducted and the proposed method achieves very promising results, especially for the JPEG images.

The remainder of this paper is organized as follows. Section \ref{sec:saak} introduces the data-driven Saak transform. Section \ref{sec:proposed} provides the detailed introduction on the proposed compressed image quality assessment method using Saak transform features. Extensive experimental results and discussions are reported in Section \ref{sec:experiments}, and finally we conclude this paper in Section \ref{sec:cons}.

\section{Saak Transform}
\label{sec:saak}

Inspired by convolutional neural networks (CNN), Kuo and Chen proposed Saak transform in \cite{kuo2018data} to well support efficient feature extraction in image understanding \cite{chen2018saak} and image reconstruction. The Saak transform consists of two main ingredients similar with traditional CNNs, \textit{i.e.}, subspace approximation and kernel augmentation. Herein, the subspace approximation utilizes data-driven kernels generated from a learning process according to the principle of energy compaction. Different from the traditional CNNs, subspace approximation in Saak utilizes the orthonormal eigenvectors of the covariance matrix of training samples as transform kernels, which are the well-known Karhunen-Lo$\grave{e}$ve transform (KLT) kernels. KLT is the optimal transform in terms of energy compaction and can provide the optimal approximation to the input with the smallest mean-squared-error (MSE) when truncating the KLT kernel functions associated with the smallest eigenvalues. To deal with various structure scales, Saak transform cascades two or more transforms directly to cover a larger receptive field.

\begin{figure}
  \centering
  \includegraphics[width=8.0cm]{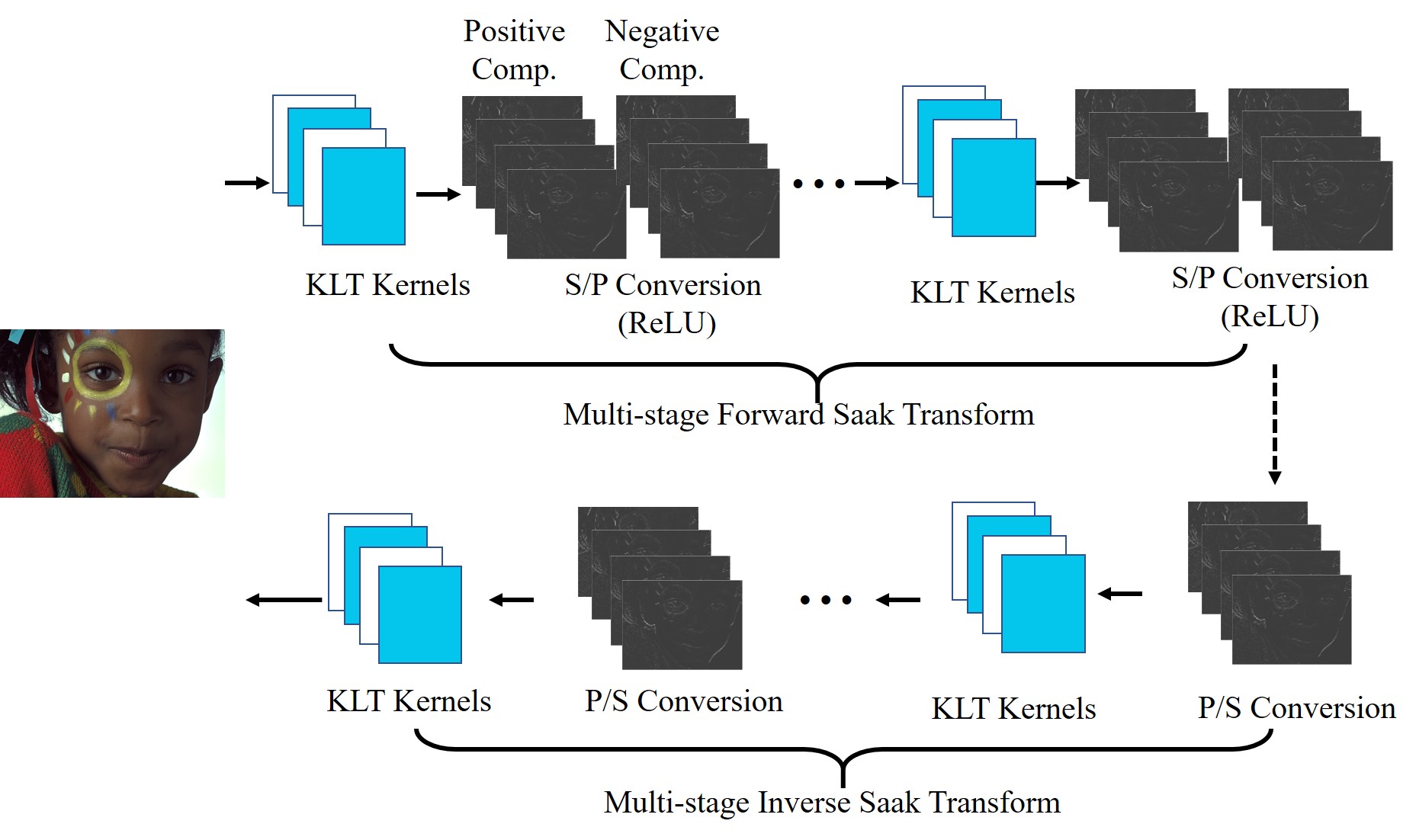}\\
  \caption{The block diagram of forward and inverse Saak transforms}\label{fig:SaakTrans}
\end{figure}

However, this cascaded operation can result in ``sign confusion'' problem \cite{kuo2017cnn}. To solve the problem, Kuo proposed to insert the Rectified Linear Unit (ReLU) activation function in between. Since the ReLU function inevitably brings up the rectification loss, Kuo further proposed the kernel augmentation strategy to eliminate this loss in Saak transform. That is, each transform kernel is augmented with its negative vector, and both original and augmented kernels are utilized in the Saak transform. When an input vector is projected onto the positive/negative kernel pair, one will go through the ReLU operation while the other will be blocked. Therefore, the Saak transform can achieve multi-stage conversion between spatial-spectral representations with lossless. Fig.\ref{fig:SaakTrans} shows the architecture of forward and inverse Saak transforms, where the ReLU operation is implemented by signal-position conversion. In forward transform, each spectral component is split into positive and negative parts via S/P conversion as shown in Eqn.(\ref{eq:SPC1}) and (\ref{eq:SPC2}), and the two parts are merged via P/S conversion in inverse transform stage.

\begin{equation}\label{eq:SPC1}
G_{k,+}(i,j) = \left\{
             \begin{array}{lr}
             G_k(i,j), & if~G_k(i,j)>0  \\
             0, & if~G_k(i,j) \leq 0,
             \end{array}
             \right.
\end{equation}

\begin{equation}\label{eq:SPC2}
G_{k,-}(i,j) = \left\{
             \begin{array}{lr}
             -G_k(i,j), & if~G_k(i,j)<0  \\
             0, & if~G_k(i,j) \geq 0.
             \end{array}
             \right.
\end{equation}
Herein, $G_{k,+}$ and $G_{k,-}$ are the positive and negative parts for the $k^{th}$ spectral component. The signed KLT coefficients in each stage are denoted as Saak coefficients. Since these coefficients reflect the dominant structural information of input images, they can serve as discriminative features of input images to carry out various image understanding tasks.

\section{Compressed Image Quality Assessment based on Saak Features}
\label{sec:proposed}
In compressed images, the main distortion source is the quantization on transformed coefficients. For JPEG images, the independent quantization for $8\times8$ blocks results in blocking artifacts, and for JPEG2000 compression, the bit plane truncation quantization for wavelet coefficients results in ringing artifacts. These two kinds of distortions are both structural distortions, and in particular HVS is more sensitive to the blocking artifacts.

\begin{figure}[t]
\begin{minipage}[b]{0.22\textwidth}
\centering
\subfigure[Monarch]{\label{fig:saaka}\includegraphics[width=3.8cm]{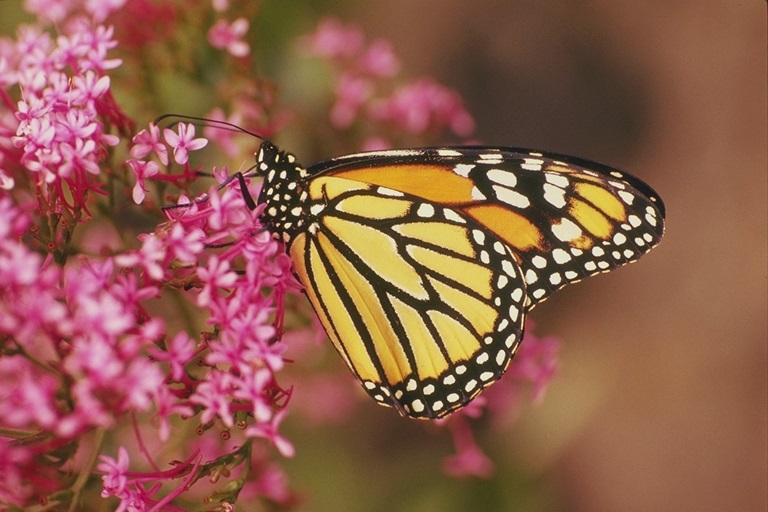} }
\end{minipage}
\hspace{0.1cm}
\begin{minipage}[b]{0.22\textwidth}
\centering
\subfigure[Energy distribution]{\label{fig:saakb}\includegraphics[width=3.7cm]{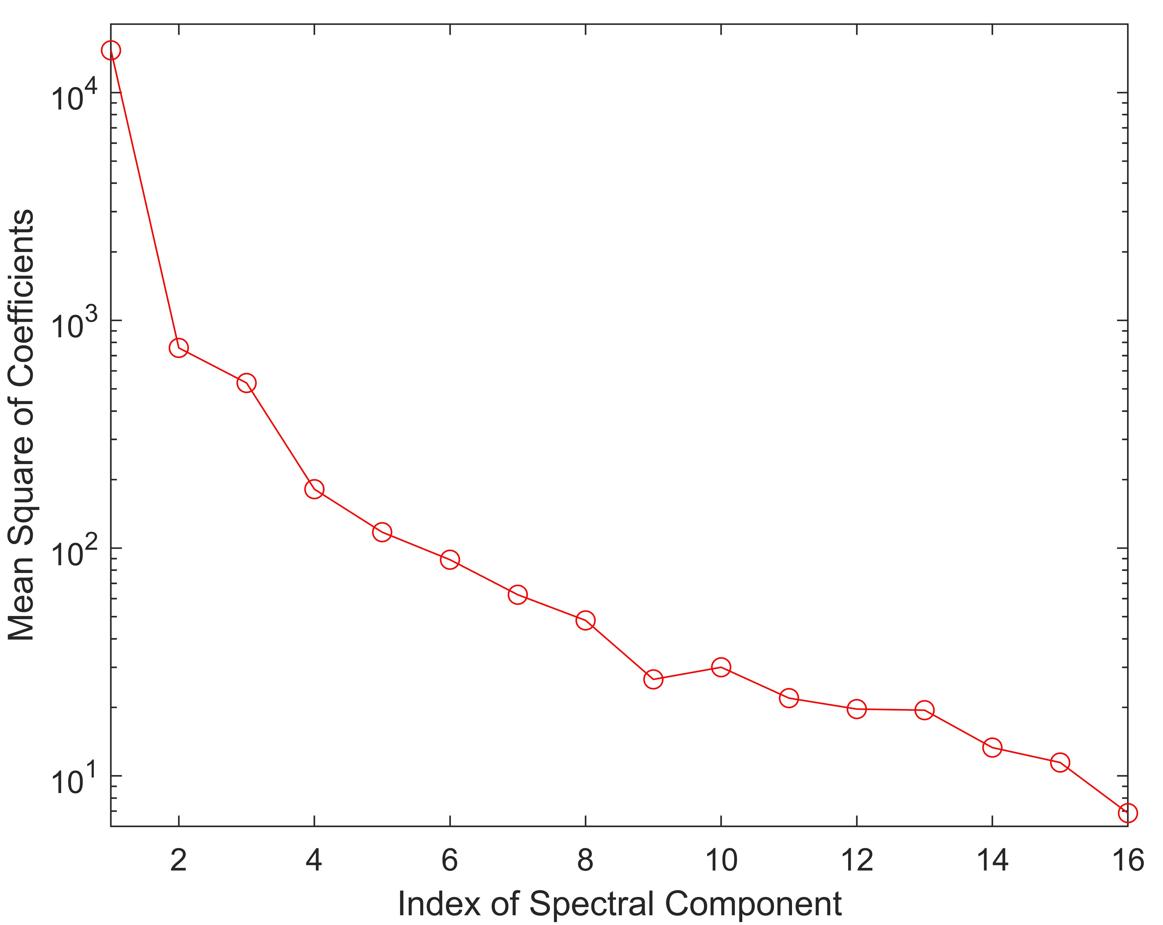} }
\end{minipage}

\begin{minipage}[b]{0.22\textwidth}
\centering
\subfigure[1$^{st}$ AC spectral component]{\label{fig:saakc}\includegraphics[width=3.8cm]{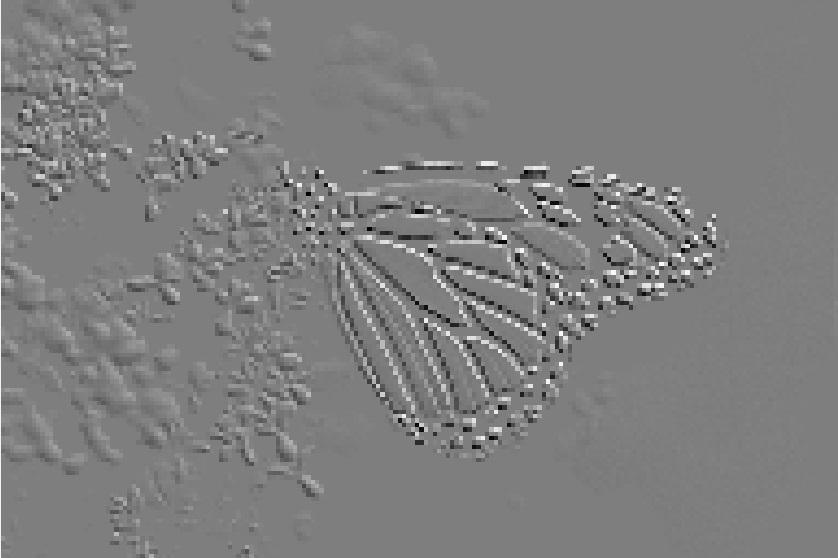} }
\end{minipage}
\hspace{0.1cm}
\begin{minipage}[b]{0.22\textwidth}
\centering
\subfigure[7$^{th}$ AC spectral component]{\label{fig:saakd}\includegraphics[width=3.8cm]{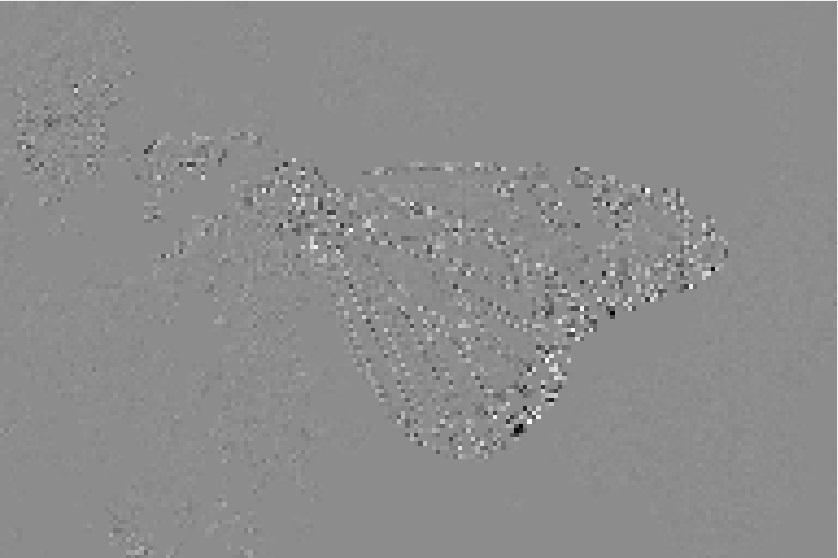} }
\end{minipage}
\caption{The Saak transformed image and the coefficient energy distribution. }
\label{fig:saakimages}
\end{figure}

Inspired by the principle of Saak transform, we utilize Saak transform as a structural information extractor to convert images into a feature domain with structural representations, to which HVS are more sensitive. Fig.\ref{fig:saakc} and Fig.\ref{fig:saakd} show the Saak coefficients in the first and $7^{th}$ AC spectral components of \textit{Monarch}, where the main structures are extracted and enhanced in transform domain especially in low frequency spectral components. In the high frequency spectral components, the Saak coefficients show very weak structures.

Since the Saak transform kernels are data-driven, their performance on feature extraction is better if the statistical characteristics of test samples are more consistent with training samples. In the proposed method, we directly extract overlapped blocks from reference image and train special Saak transform matrix, $\mathbb{T}$, for individual image content. To avoid the influence of noisy blocks without meaningful structures, we only utilize the blocks with their pixel standard deviation larger than 2 as training samples.

To calculate the objective quality score for one distorted image, we first apply a low-pass Gaussian filter, $\mathbb{G}$, to both the reference and distorted images to remove the inevitable noisy information, which is usually masked by HVS. Then, the learnt Saak transform from the corresponding reference image is utilized to convert the reference and distorted images into spectral domain. Finally, a quality function is formulated based on these Saak coefficients, which is expressed as follows,
\begin{equation}
s = f(\mathbb{T}(\mathbb{G}\ast I_{org})),\mathbb{T}(\mathbb{G}\ast I_{dist})).
\label{eq:quality}
\end{equation}

There are two popular kinds of quality functions, \textit{i.e.,} MSE and correlation based functions, which are utilized in PSNR and SSIM based quality metrics. Considering their advantages in different distortion types, we design our quality function $f(\cdot)$ as the weighted combination of the two kinds functions,
\begin{equation}
f(F_{ref}, F_{dist}) = (1-\lambda)e^{\frac{-\sum_{k=1}^{K}w_k D_k}{c}} + \lambda \sum_{k=1}^{K}w_k C_k,
\label{eq:proposed}
\end{equation}
where $F_{ref}$ and $F_{dist}$ are the transformed reference and distorted images respectively, $D_k$ is the MSE between $F_{ref,k}$ and $F_{dist,k}$, which correspond to the $k^{th}$ spectral components of reference and distorted images. $C_k$ is the correlation coefficient of $F_{ref,k}$ and $F_{dist,k}$. Here, $K$ is the number of spectral components, $\lambda$ is a constant factor and $w_k$ is a weighting factor.

Except for the augmentation, the Saak transform kernels actually are the same with KLT kernels, which can well concentrate the image energy in the low frequency spectral components. As shown in Fig.\ref{fig:saakb}, we can see the spectral component energy decreases rapidly. By joint analyzing Fig.\ref{fig:saakc} and Fig.\ref{fig:saakd}, we can assume that the dominant image structures are extracted into the low frequency spectral components, while non-structural information is decomposed into the high frequency spectral components. The quality of low frequency spectral components plays more important role in the compressed image quality assessment problem. Therefore, we proposed a weighting function according to the energy distribution of spectral components,
\begin{equation}
w_k = \frac{1}{Z} (1 - e^{-\frac{E_k}{h^2}}),
\label{eq:weighting}
\end{equation}
where $E_k$ is the mean square of the coefficients in $F_{ref,k}$ and $F_{dist,k}$, and $h$ is a constant parameter. $Z$ is a normalization factor.

\section{Experimental Results and Analyses}
\label{sec:experiments}
In this section, we evaluate the efficiency of the proposed method on two popular IQA datasets, LIVE \cite{LIVEdataset} and CSIQ \cite{larson2010most}, and compare our method with some state-of-the-art IQA algorithms, including IFC \cite{sheikh2005information}, IWSSIM \cite{wang2011information}, MAD \cite{larson2010most}, MS-SSIM \cite{wang2003multiscale}, PSNR, PSNR-HVS \cite{egiazarian2006new}, RFSIM \cite{zhang2010rfsim}, SSIM \cite{wang2004image}, UQI \cite{wang2002universal}, VIF \cite{sheikh2004image}, and VSI \cite{zhang2014vsi}. There are 30 reference images in LIVE and CSIQ respectively, but the reference images in the two datasets are different contents without duplication. There are two kinds of compressed images generated by JPEG and JPEG2000 respectively with different distortion levels in these datasets. To evaluate the performance of these IQA methods, we utilized three widely utilized correlation coefficients, \textit{i.e.,} Pearson linear correlation coefficient (PLCC), Spearman rank-order correlation coefficient (SRCC), Kendall rank-order correlation coefficient (KRCC). The objective scores from a better IQA method should have higher correlation coefficients with subjective scores. Herein, the PLCC metric is calculated between MOS and the objective scores after nonlinear regression, and the widely used nonlinear regression function proposed in \cite{sheikh2006statistical} is as follows,
\begin{equation}
q(x) = \beta_1\left(\frac{1}{2} - \frac{1}{1+e^{\beta_2(x-\beta_3)}}\right)+\beta_4 x + \beta_5.
\label{eq:regression}
\end{equation}

\begin{table*}[htbp]
  \centering
  \renewcommand{\arraystretch}{1.15}
  \newcommand{\tabincell}[2]{\begin{tabular}{@{}#1@{}}#2\end{tabular}}
  \caption{The PLCC, SRCC and KRCC between the subjective scores and the objective scores from different IQA methods on LIVE and CSIQ datasets.}
    \begin{tabular}{p{1.1cm}|p{0.9cm}<{\centering}|p{0.9cm}<{\centering}|p{0.9cm}<{\centering}|p{0.9cm}<{\centering}|p{0.9cm}<{\centering}|p{0.9cm}<{\centering}|p{0.9cm}<{\centering}|p{0.9cm}<{\centering}|p{0.9cm}<{\centering}|p{0.9cm}<{\centering}|p{0.9cm}<{\centering}|p{0.9cm}<{\centering} }
     \hline
\multirow{3}{*}{\tabincell{c}{IQA \\ methods}} &\multicolumn{6}{c|}{LIVE} &\multicolumn{6}{c}{CSIQ} \\ \cline{2-13}
             &\multicolumn{3}{c|}{JPEG} &\multicolumn{3}{c|}{JPEG 2000} &\multicolumn{3}{c|}{JPEG} &\multicolumn{3}{c}{JPEG 2000} \\ \cline{2-13}
             &PLCC  &SRCC  &KRCC  &PLCC  &SRCC  &KRCC &PLCC  &SRCC  &KRCC  &PLCC  &SRCC  &KRCC\\\hline
IFC        &0.8992	&0.8661	&0.6756	&0.9022	&0.8920	&0.6967	&0.95	&0.9412	&0.7975	&0.9331	&0.9252	&0.781\\\hline
IWSSIM     &0.9412	&0.9074	&0.7441	&0.9578	&0.9501	&0.8001	&\textbf{0.9844}	&0.9662	&\textbf{0.8421}	&0.9809	&0.9683	&0.848\\\hline
MAD        &0.9384	&0.9055	&0.7375	&\textbf{0.9618}	&\textbf{0.9531}	&\textbf{0.8114}	&0.9827	&0.9615	&0.8351	&\textbf{0.9836}	&\textbf{0.9752}	&\textbf{0.8709}\\\hline
MSSIM      &0.9429	&\textbf{0.9131}	&\textbf{0.7487}	&0.9572	&0.9529	&0.8042	&0.9823	&0.9634	&0.8332	&0.9777	&0.9684	&0.8439\\\hline
PSNR       &0.8596	&0.8409	&0.6359	&0.8964	&0.8898	&0.7037	&0.8799	&0.8881	&0.6936	&0.9451	&0.9362	&0.7667\\\hline
\multicolumn{1}{m{1.2cm}|}{PSNR-HVS}   &0.9352	&0.9040	&0.7215	&0.9521	&0.9454	&0.7932	&0.9697	&0.9514	&0.8032	&0.9769	&0.9703	&0.8449\\\hline
RFSIM      &0.9255	&0.8930	&0.7084	&0.9350	&0.9286	&0.7635	&0.9617	&0.9481	&0.7954	&0.9667	&0.9635	&0.8306\\\hline
SSIM       &0.9297	&0.9028	&0.7149	&0.9368	&0.9317	&0.7633	&0.942	&0.9222	&0.7545	&0.9236	&0.9207	&0.754\\\hline
UQI        &0.8518	&0.8273	&0.6216	&0.8487	&0.8466	&0.6434	&0.9182	&0.9078	&0.728	&0.9043	&0.8813	&0.7088\\\hline
VIF        &0.9302	&0.9047	&0.7326	&0.9576	&0.9524	&0.8026	&0.9228	&\textbf{0.9684}	&0.8416	&0.9782	&0.9697	&0.8501\\\hline
VSI        &\textbf{0.9443}	&0.9089	&0.7308	&0.9533	&0.9480	&0.7944	&0.9808	&0.9618	&0.8271	&0.9745	&0.9694	&0.8482\\\hline
Proposed   &\textbf{0.9441}	&\textbf{0.914}	&\textbf{0.7517}	&\textbf{0.9625}	&\textbf{0.9542}	&\textbf{0.809}	&\textbf{0.9878}	&\textbf{0.9691}	&\textbf{0.8489}	&\textbf{0.9813}	&\textbf{0.9737}	&\textbf{0.861}\\\hline
    \end{tabular}%
  \label{tab:CC}%
\end{table*}

\begin{figure}[t]
\begin{minipage}[b]{0.22\textwidth}
\centering
\subfigure[JPEG images]{\label{fig:curve_a}\includegraphics[width=3.6cm]{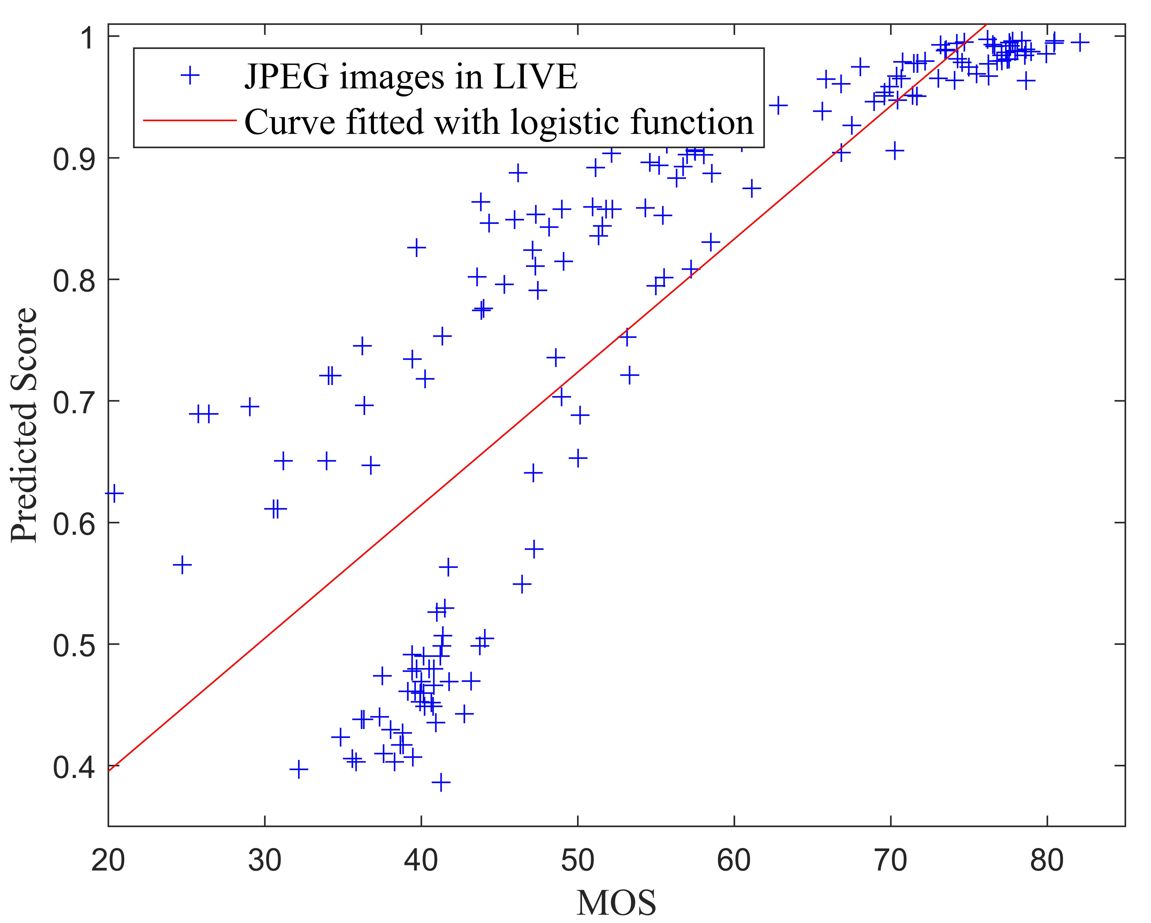} }
\end{minipage}
\hspace{0.1cm}
\begin{minipage}[b]{0.22\textwidth}
\centering
\subfigure[JPEG2000 images]{\label{fig:curve_b}\includegraphics[width=3.6cm]{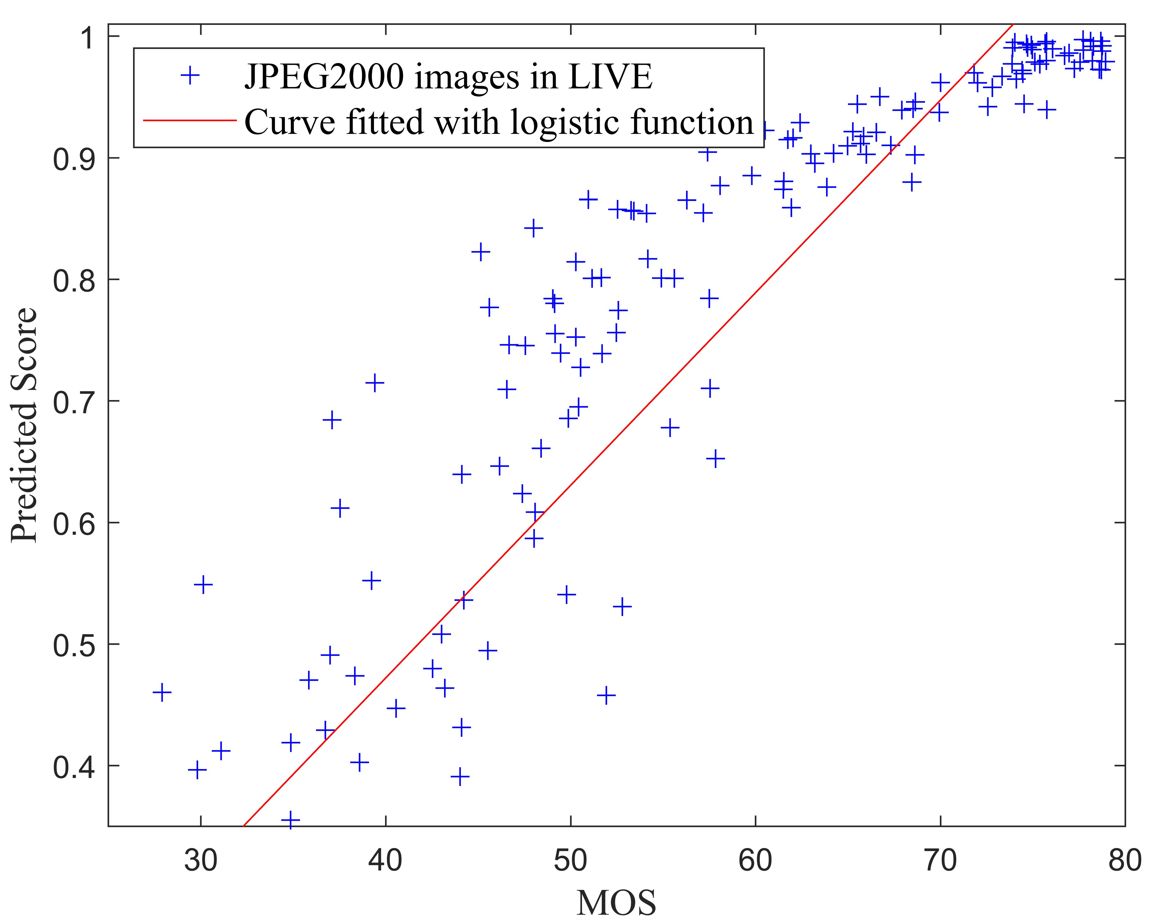} }
\end{minipage}
\caption{Scatter plots of subjective scores versus the predicted scores from the proposed method on LIVE dataset (a) JPEG, (b) JPEG2000. }
\label{fig:curve}
\end{figure}

In our experiment, the parameter $c$ in Eqn.(\ref{eq:proposed}) is set 400, and $h$ in Eqn.(\ref{eq:weighting}) is set 100. Since JPEG and JPEG2000 are different compression frameworks, which are suitable for different quality functions, we set the parameter $\lambda$ as 0.7 for JPEG images and 0.2 for JPEG2000 images. In our experiments, the two-stage Saak transform with $4\times 4$ block size is utilized, and each image is transformed into 496 spectral components after P/S conversion, \textit{i.e.,} $K=496$. Table \ref{tab:CC} lists all the three correlation coefficients on the two datasets, where the top2 performances are highlighted by boldface. We can see that the proposed method achieves the best performance on the two datasets based on the three correlation coefficients. In particular, the proposed method achieves much better performance on JPEG compressed images compared with that on JPEG2000 images. This is because in JPEG images, the blocking artifacts are more destructive to image structures compared with the ringing artifacts in JPEG2000 images. Although MAD achieves better quality assessment performance on CSIQ JPEG2000 images, it is obviously inferior to the proposed method on the other cases. The proposed method shows strong robustness on different datasets which verifies that the structural information extracted by Saak transform is more meaningful for HVS.

In Fig.\ref{fig:curve}, we illustrate the scatter plots of the subjective scores and the predicted scores from the proposed method on LIVE dataset. Herein, the curves shown in Fig.\ref{fig:curve} are obtained by the nonlinear fitting function in Eqn.(\ref{eq:regression}). We can see that the proposed method provides very consistent results with the subjective ones.

\section{Conclusion}
\label{sec:cons}
In this paper, we proposed a new compressed image quality assessment method utilizing the Saak transform to extract dominant features. The proposed method measures the compressed image quality in Saak transform domain according to feature distortions instead of pixel-level distortions in spatial domain. The feature distortions in different spectral components are weighted according to the feature importance which is measured by the energy of spectral components in Saak transform domain. Finally, the MSE and correlation based quality functions are combined to predict the final objective quality score. Experimental results on two popular IQA datasets show that the proposed method achieved better performance and strong robustness compared with the widely utilized IQA methods. This work also illustrated the efficiency of the Saak transform in image structural information representation.

\bibliographystyle{IEEEbib}
\bibliography{SaakTransformBasedCompressedImageQualityAssessment}

\end{document}